%% file: main_text.tex
\documentclass[10pt]{article}

\usepackage{cite}
\usepackage{authblk} 
\usepackage[a4paper, margin=0.8in]{geometry}
\usepackage{graphicx}
\usepackage{amsmath}
\usepackage{booktabs}
\usepackage{multirow}
\usepackage{geometry}
\usepackage{caption}
\usepackage{booktabs}
\usepackage{tabularx}
\usepackage{ulem}
\usepackage{soul}
\usepackage{xcolor}
\sethlcolor{yellow}
\usepackage{placeins} 
\usepackage{float}

\title{\Large \textbf{Direction-Dependent Conduction Polarity in Altermagnetic CrSb}}
\author[1,3]{Banik Rai}
\author[1,3]{Krishnendu Patra}
\author[2]{Satyabrata Bera}
\author[2]{Sk Kalimuddin}
\author[1]{Kakan Deb}
\author[2]{Mintu Mondal}
\author[1,*]{Priya Mahadevan}
\author[1,**]{Nitesh Kumar}

\affil[1]{Department of Condensed Matter and Materials Physics, S. N. Bose National Centre for Basic Sciences, Salt Lake City, Kolkata-700106, India}
\affil[2]{School of Physical Sciences, Indian Association for the Cultivation of Science, Jadavpur, Kolkata-700032, India}
\affil[3]{These authors contributed equally to this work.}
\affil[*]{Contact author: priya@bose.res.in}
\affil[**]{Contact author: nitesh.kumar@bose.res.in}

\date{}

\begin{document}
\maketitle

\textbf{CrSb has recently gained immense attention as an altermagnetic candidate. This work reports on the experimental observation of direction-dependent conduction polarity (DDCP) in altermagnetic CrSb through Hall and Seebeck thermopower measurements. Conduction is dominated by holes along the \textit{c}-axis and by electrons in the \textit{ab}-plane of the hexagonal crystal of CrSb. Density functional theory (DFT) calculations indicate that DDCP in CrSb arises from a multicarrier mechanism, where electrons and holes living in distinct bands dominate conduction along different crystallographic directions. Furthermore, DFT predicts that DDCP exists within a narrow energy window near the Fermi level and is sensitive to small doping levels. This prediction is experimentally validated by the loss of DDCP in hole-doped Cr$_{0.98}$V$_{0.02}$Sb. These findings highlight the potential for tunable electronic behavior in CrSb, offering promising avenues for applications in devices that require both p-type and n-type functionalities within a single material.}

\maketitle

\section{Introduction}

\begin{figure}[h]
\centering
\includegraphics[width=0.5\linewidth]{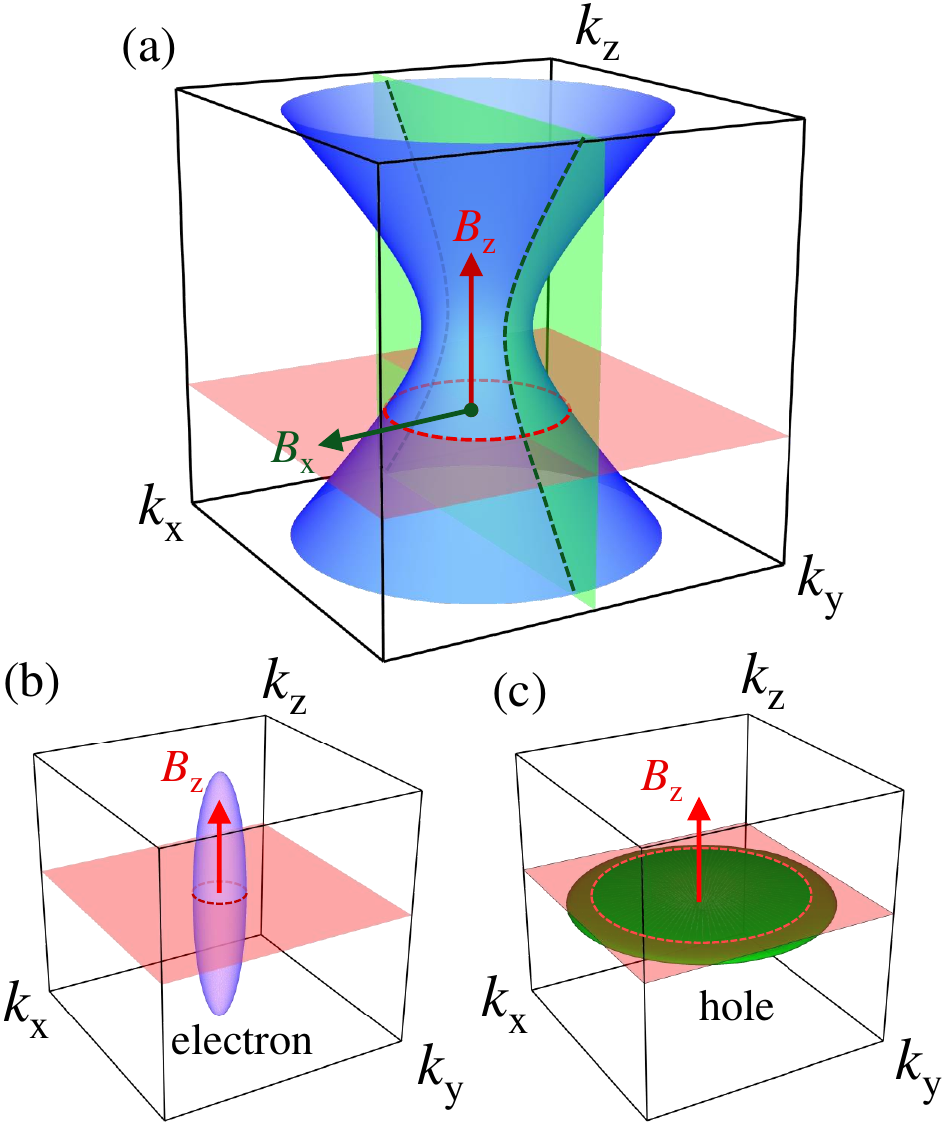}
\caption{\label{fig1} FS geometries for DDCP via single carrier and multicarrier mechanisms. (a) Hyperboloid FS for single carrier mechanism exhibiting convex character in the $k_\mathrm{x}-k_\mathrm{y}$ plane and concave character along the $k_\mathrm{z}$ direction, with dashed lines representing the possible charge carrier trajectories for different magnetic field directions. (b)-(c) Anisotropic Fermi pockets for the multicarrier mechanism, where, without loss of generality, the left pocket represents an electron pocket and the right pocket represents a hole pocket. The electron pocket is elongated along the $k_\mathrm{z}$ direction, while the hole pocket is flat in the $k_\mathrm{x}-k_\mathrm{y}$ plane. A possible electron and hole trajectory is shown with dashed lines for a magnetic field applied along the z-direction. Electrons dominate conduction in the $k_\mathrm{x}-k_\mathrm{y}$ plane, while the holes dominate conduction along $k_z$-direction.}
\end{figure}

Almost all materials (metals and semiconductors) have a single type of dominant charge carrier, either electrons or holes, along all crystallographic directions. This allows these materials to be easily classified as either n-type (electron dominant) or p-type (hole dominant). However, there exists a rare and distinct class of materials that defies this simple classification.\cite{He2019,Koster2023,Nakamura2021,Helman2021,Ochs2021,Wang2020,Qi2022} These unique materials can exhibit electron- and hole-dominant conduction simultaneously along different crystallographic directions. This direction-dependent conduction polarity (DDCP), is enabled by the geometrical features of the Fermi surface (FS). The effective mass tensor of a charge carrier moving through a crystal lattice potential is given by:
\begin{eqnarray}
\label{eq1}
m_{ij}^*=\hbar^2\left[\frac{\partial^2E}{{\partial k_i}{\partial k_j}}\right]^{-1}_{E=E_\mathrm{F}}.
\end{eqnarray}
Electrons have positive $m_{ij}^*$ while holes have negative $m_{ij}^*$. Since the second derivative of a function determines its curvature, $m_{ij}^*$ can be positive or negative depending on the curvature of the corresponding energy band crossing the Fermi energy ($E_\mathrm{F}$). DDCP can be realized if the integrated curvature of all the bands crossing $E_\mathrm{F}$ is positive in one direction and negative in the other. The FS, which provides a geometric representation of all points in reciprocal space where the band energy equals $E_\mathrm{F}$, encompasses all bands crossing $E_\mathrm{F}$. A single band material can exhibit DDCP if its FS has concave and convex characters along different directions (see Figure \ref{fig1}a). This has been observed in $\mathrm{NaSn_2As_2}$,\cite{He2019} which has a hyperboloid-shaped FS that has both positive and negative curvatures (dependent on direction). The DDCP arising from single carrier mechanism is termed ``goniopolarity", and the materials exhibiting it are called ``goniopolar". Another way to achieve DDCP is through multicarrier mechanism (see Figure \ref{fig1}(b-c)), where electrons and holes live in distinct bands and dominate conduction along different crystallographic directions.\cite{Koster2023, Wang2020} This can be achieved when the FS exhibits highly anisotropic electron and hole pockets, such that the integrated curvature of the total FS has opposite signs along two different crystallographic directions.

Although materials exhibiting DDCP have recently gained attention as novel thermoelectric materials,\cite{Tang2015, Manako2024, Scudder2021} only a handful of such materials have been discovered to date. Among these, very few are both air-stable and composed of earth-abundant elements, which are essential prerequisites for practical device applications and commercialization. Additionally, inconsistencies often arise in the experimental characterization of reported DDCP materials. For example, the Hall coefficient and Seebeck coefficient, whose signs are key indicators of DDCP, may exhibit conflicting signs \cite{He2019} or one of them fail to demonstrate the direction-dependent opposite signs,\cite{Manako2024} which is the defining feature of DDCP.

In this paper, we investigate the transport properties of the recently identified altermagnetic candidate CrSb. Altermagnets represent a novel phase of collinear magnets with vanishing magnetization and spin-split energy bands,\cite{vsmejkal2022beyond, smejkal2022} making them an important material for future spintronics applications.  We report the observation of DDCP in CrSb, based on Hall and thermopower measurements. Both the Hall coefficient and Seebeck coefficient indicate that the conduction is dominated by holes along the \textit{c}-axis and by electrons in the \textit{ab}-plane of the CrSb hexagonal crystal. This behavior is qualitatively supported by our DFT calculations. The FS calculated by DFT consists of two hole pockets and one electron pocket. The DDCP in CrSb emerges as the electron and hole pockets dominate conduction along different crystallographic directions. DFT also predicts that the DDCP is confined in a narrow energy window ($\Delta E \approx 14\: \mathrm{meV}$) near $E_\mathrm{F}$ and can be suppressed by doping. This prediction is experimentally validated, as a doping of $\sim 0.04$ number of holes per unit cell in Cr$_{0.98}$V$_{0.02}$Sb destroys the DDCP and drives the system towards p-type.

Given that both Cr and Sb are earth-abundant and relatively non-toxic elements, CrSb holds great promise for practical applications that exploit its DDCP behavior. Furthermore, its altermagnetic properties introduce an intriguing aspect, paving the way for future spintronics applications that leverage both DDCP and spin-splitting phenomena.

\begin{figure}[h]
\centering
\includegraphics[width=0.7\linewidth]{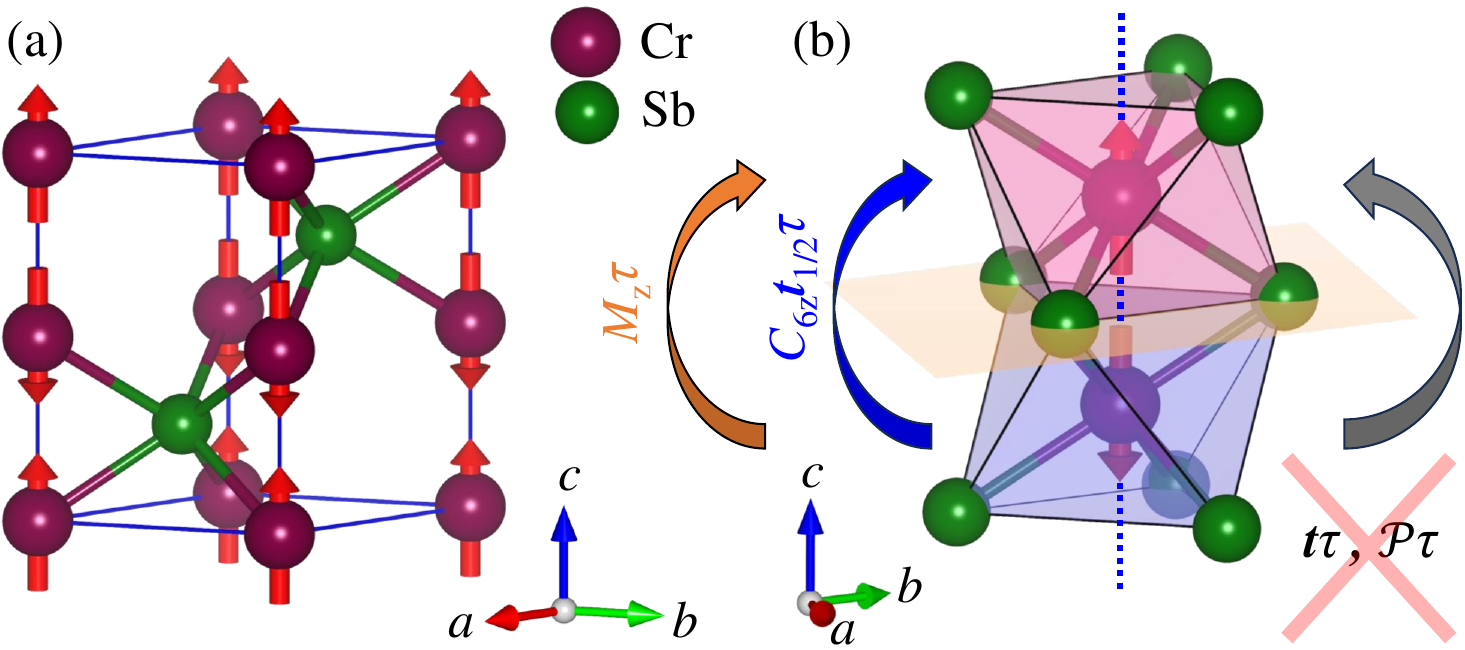}
\caption{\label{fig2} Crystal structure and sublattice transposing symmetries of CrSb. (a) Magnetic unit cell of CrSb showing A-type antiferromagnetic ordering of Cr-moments. (b) Sublattice transposing symmetries of CrSb showing that a lattice translation or inversion does not connect the opposite spin sublattices, which are instead connected by a horizontal mirror passing through Sb layer or a real space crystal rotation combined with half the lattice translation along the \textit{c}-axis.}
\end{figure}

\section{Results and Discussion}
CrSb crystallizes in a hexagonal NiAs-type structure with space group 
$P6_3/mmc$ and exhibits A-type antiferromagnetic order with a Néel temperature of approximately 703 K.\cite{Snow1952,Yuan2020} The Cr moments are aligned parallel to the \textit{c}-axis, coupled ferromagnetically in the \textit{ab}-plane and antiferromagnetically along the \textit{c}-axis [see Figure \ref{fig2}a].  The anisotropic arrangement of Sb atoms around the oppositely aligned Cr spins prevents the connection of opposite spin sublattices by simple lattice translations or inversion (see Figure \ref{fig2}b). However, these sublattices can be connected either by a six-fold crystal rotation about the \textit{c}-axis $(C_\mathrm{6z})$ combined with half the lattice translation along the \textit{c}-axis $(\textit{\textbf{t}}_{1/2} = (0, 0, c/2))$ or by a horizontal mirror $(M_\mathrm{z})$ passing through the Sb layer. These operations, when combined with time-reversal symmetry ($\tau$), yield the actual sublattice-transposing symmetries of the crystal: $C_\mathrm{6z}\textit{\textbf{t}}_{1/2}\tau$ and $M_\mathrm{z}\tau$. These symmetries give rise to four spin-degenerate nodal planes in the Brillouin zone (BZ) (see Figure S6a). Away from these high-symmetry nodal planes, the spin degeneracy is lifted, resulting in significant non-relativistic splitting of the energy bands (see Figure S6b).

\begin{figure*}[t]
\centering
\includegraphics[width=0.95\linewidth]{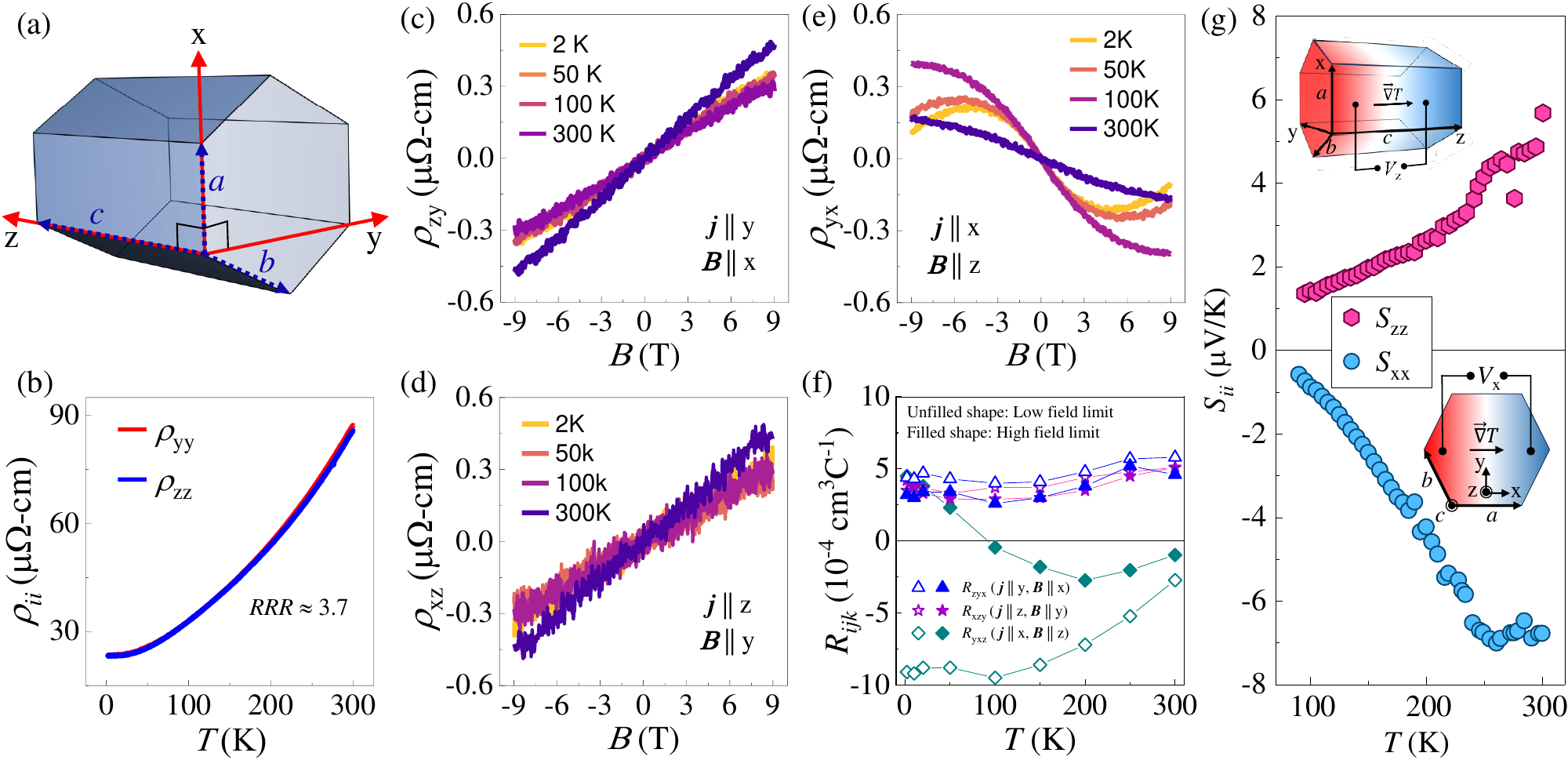}
\caption{\label{fig3} Electrical transport properties and Seebeck effect in CrSb. (a) Defining coordinate axes on the hexagonal unit call of CrSb. (b) In-plane $(\rho_\mathrm{yy})$ and cross-plane $(\rho_\mathrm{zz})$ longitudinal resistivity as a function of temperature. (c)-(e) Magnetic field dependent Hall resistivity for different Hall geometries. (f) Variation of Hall coefficient as a function of temperature for different Hall geometries. (g) In plane $(S_\mathrm{xx})$ and cross-plane $(S_\mathrm{zz})$ Seebeck coefficients as a function of temperature.    }
\end{figure*}

To examine the transport behavior of CrSb along different crystallographic directions, the coordinate axes have been defined on the hexagonal lattice of CrSb (see Figure \ref{fig3}a). The x- and z-axes align with the crystallographic \textit{a}- and \textit{c}-directions whereas the y-axis is along the crystallographic $[01\bar{1}0]$ direction. Figure \ref{fig3}b shows the temperature dependence of in-plane $(\rho_\mathrm{yy})$ and cross-plane $(\rho_\mathrm{zz})$ longitudinal resistivity. The material shows metallic nature throughout the temperature range in both in-plane and cross-plane directions with almost coinciding resistivity values and similar residual resistivity ratios $(RRR \approx3.7)$. The field dependent Hall resistivity data in different geometries are shown in Figure \ref{fig3}(c-e). When a magnetic field is applied within the hexagonal plane of the CrSb crystal, as illustrated in Figure \ref{fig3}c and \ref{fig3}d for $\textit{B}||\mathrm{x}$ and $\textit{B}||\mathrm{y}$ respectively, both the Hall resistivity $\rho_\mathrm{zy}$ and $\rho_\mathrm{xz}$ exhibit an almost linear dependence on the field at all temperatures with similar positive slopes. When the magnetic field is applied along the z-axis, the Hall resistivity $\rho_\mathrm{yx}$ deviates from linearity, especially at lower temperatures, and the slopes becomes predominantly negative. This non-linear behavior in Hall resistivity is often associated with the anomalous Hall effect \cite{Nagaosa2010} or a multicarrier transport mechanism.\cite{hurd2012hall} Since the orientation of the Néel vector in CrSb precludes the occurrence of anomalous Hall effect,\cite{Smolyanyuk2024,Zhou2025,Urata2024,yu2024n} the non-linear  behavior can be attributed to the multicarrier mechanism (see Supporting Information).\cite{Urata2024, bai2024nonlinear} Figure \ref{fig3}f presents the Hall coefficient, $R_{ijk}$, as a function of temperature for all Hall measurement geometries, evaluated at 0 T (low-field limit) and 9 T (high-field limit) using the equation: $R_{ijk}\left(B_0\right)=\left.\dfrac{d\rho_{ij}}{d B_k}\right|_{B_0}$. In a multicarrier system, the non-linear nature of Hall resistivity results in distinct low-field and high-field values for the Hall coefficient. The low-field value carries information of both carrier density and mobility, while the high-field value only carries the information of carrier denisty. \cite{hurd2012hall, ashcroft1978solid}  Since both parameters play a crucial role in the overall conduction, the low-field value of the Hall coefficient is more reliable for determining the dominant conduction type in a material, with positive value indicating hole-dominant conduction and negative value indicating electron-dominant conduction. Across all temperatures, the low-field value of $R_\mathrm{zyx}$ and $R_\mathrm{xzy}$ remain positive, while that of $R_\mathrm{yxz}$ remains negative, inferring that the transport is dominated by electrons when the magnetic field is applied along the \textit{c}-axis and by holes when the magnetic field is applied in the \textit{ab}-plane of the hexagonal CrSb crystal. The sign of $R_{ijk}$ is found to depend on whether the magnetic field is applied in-plane or out-of-plane rather than on the direction of the applied current. In particular, the trajectory of charge carrier is given by the intersection of the FS by a plane perpendicular to the magnetic field as shown in Figure \ref{fig1}.\cite{hurd2012hall, ashcroft1978solid} The applied electric field (i.e; current) induces only a minute distortion in the FS and thus negligibly affects the trajectory of the charge carriers. Consequently, an in-plane magnetic field induces cross-plane conduction, whereas a cross-plane magnetic field induces in-plane conduction. This suggests that the in-plane conduction is electron-dominated whereas the cross-plane conduction is hole dominated. The identical sign and magnitude of $R_\mathrm{zyx}$ and $R_\mathrm{xzy}$ is a direct consequence of the magnetic point group $(6'/m'mm')$ symmetry of CrSb which imposes the constraint $R_\mathrm{zyx}=R_\mathrm{xzy}$.\cite{birss1964symmetry,Urata2024}

An alternative and potentially more reliable tool for determining the polarity of charge carriers in a material is the thermopower experiment, as it does not involve the application of an external magnetic field. By applying a temperature gradient across the material and measuring the longitudinal voltage generated, the Seebeck coefficient $(S_{ii})$, which quantifies the voltage generated per unit temperature difference $S_{ii}=-\dfrac{\Delta{V}_i}{\Delta{T}_i}$, can be defined.\cite{behnia2015fundamentals} Electrons and holes produce negative and positive Seebeck coefficients, respectively. Thus, by examining the sign of Seebeck coefficients measured along different crystallographic directions, the type of charge carrier dominating the conduction along those directions can be determined. Figure \ref{fig3}g shows the temperature dependence of both the in-plane $(S_\mathrm{xx})$ and cross-plane $(S_\mathrm{zz})$ Seebeck coefficients. In the measured temperature range, $S_\mathrm{xx}$ increases towards negative values, while $S_\mathrm{zz}$ increases towards positive values with temperature. Notably, $S_\mathrm{xx}$ remains negative while $S_\mathrm{zz}$ remains positive at all measured temperatures, indicating electron-dominant conduction in the plane and hole-dominant conduction across the plane of the hexagonal CrSb crystal. Although electrons and holes produce Seebeck coefficients with opposite signs, the opposite signs of the Seebeck coefficient can, in principle, arise from other direction-dependent terms in the following tensor expression of the Seebeck coefficient: \cite{He2019,Wang2020}
\begin{eqnarray}
\label{eq2}
S_{ii}=-\frac{\pi^2K_B^2T}{3\left|e\right|}\left[\frac{1}{n(E)}\frac{dn(E)}{dE}+\frac{1}{\tau_{ii}(E)}\frac{d\tau_{ii}}{dE}+m_{ii}^\ast\frac{d}{dE}\left(\frac{1}{m_{ii}^\ast}\right)\right]_{E=E_\mathrm{F}}.
\end{eqnarray}
Here, $n(E)$ and $\tau_{ii}(E)$ are the energy dependent carrier concentration and relaxation-time tensor. Although $n(E)$ is a scalar and cannot cause anisotropy in the sign of $S_{ii}$, $\tau_{ii}(E)$ is a tensor and can have different values in different directions. However, the quantity $\dfrac{1}{{\tau_{ii}(E)}}\dfrac{d\tau_{ii}}{dE}$, which appears in the expression of $S_{ii}$, depends on the nature of the scattering mechanism of the conduction electrons and must have an isotropic sign in a clean single crystal having negligible defects. The isotropy of scattering processes in CrSb is also evidenced by the isotropic values of in-plane and cross-plane longitudinal resistivity. Thus, the anisotropy in the sign of $S_{ii}$ must arise from $m_{ii}^\ast$ as the quantity $\dfrac{d}{dE}\left(\dfrac{1}{m_{ii}^\ast}\right)$ is always negative for energy bands with cosine-like dispersion.\cite{He2019}

\begin{figure*}[!t]
\centering
\includegraphics[width=0.8\linewidth]{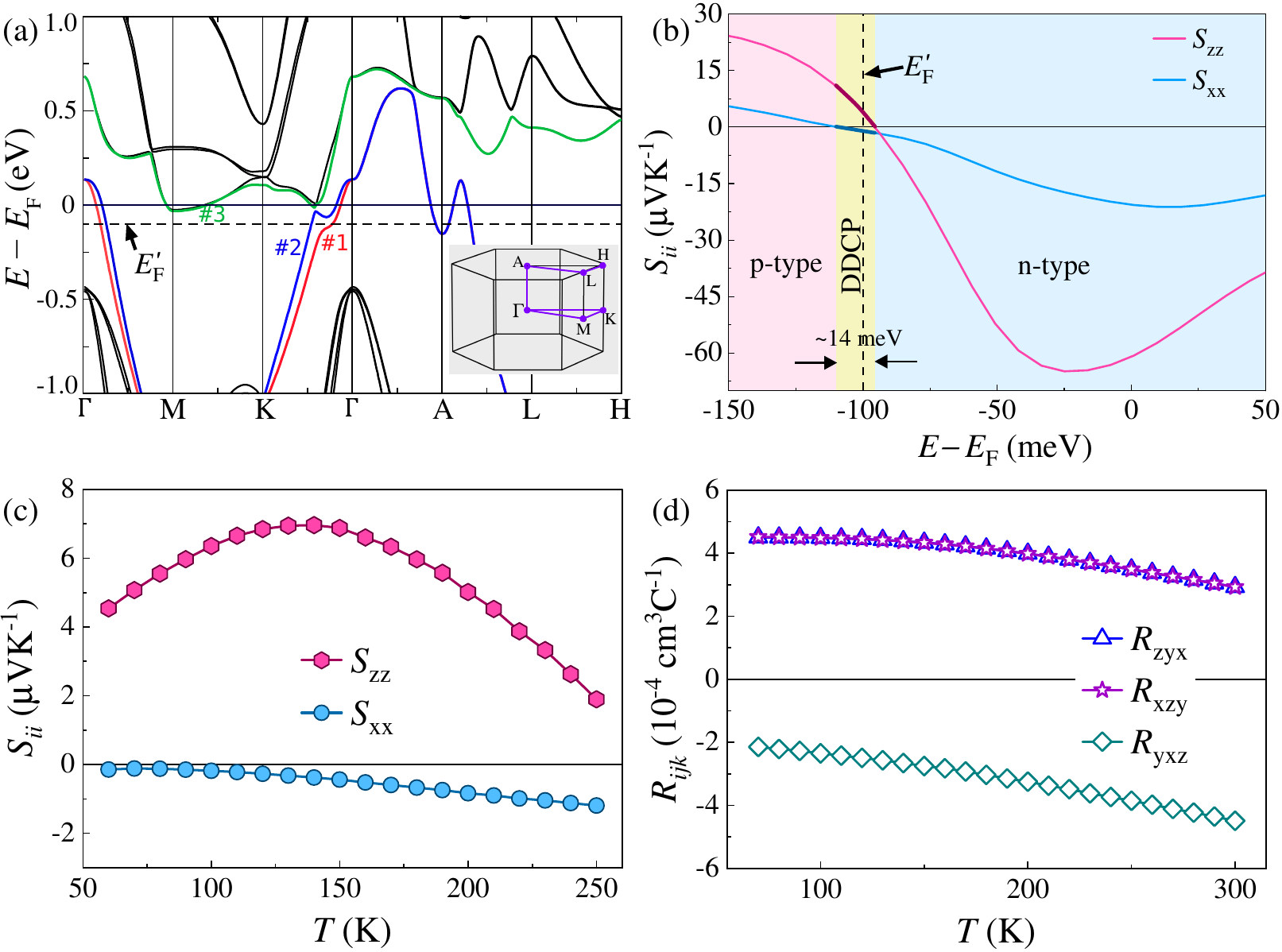}
\caption{ Results of theoretical calculations on CrSb. a Band structure along high-symmetry paths (see inset) of the BZ with SOC turned on. The bands crossing the shifted $E'_\mathrm{F}$ (dashed line) are highlighted in color and labeled numerically.\ (b) In-plane ($S_\mathrm{xx}$) and cross-plane ($S_\mathrm{zz}$) Seebeck coefficients at 100 K as a function of change in Fermi energy showing the DDCP behavior existing within a small energy window. The black dashed line represents the shifted Fermi energy $E'_\mathrm{F}$ (see text). (c) Temperature dependence of $S_\mathrm{xx}$ and $S_\mathrm{zz}$ calculated at $E'_\mathrm{F}$. (d) Temperature dependence of the Hall coefficient components calculated at $E'_\mathrm{F}$.}
\label{fig4}
\end{figure*}

\begin{figure}[h]
\centering
\includegraphics[width=0.6\columnwidth]{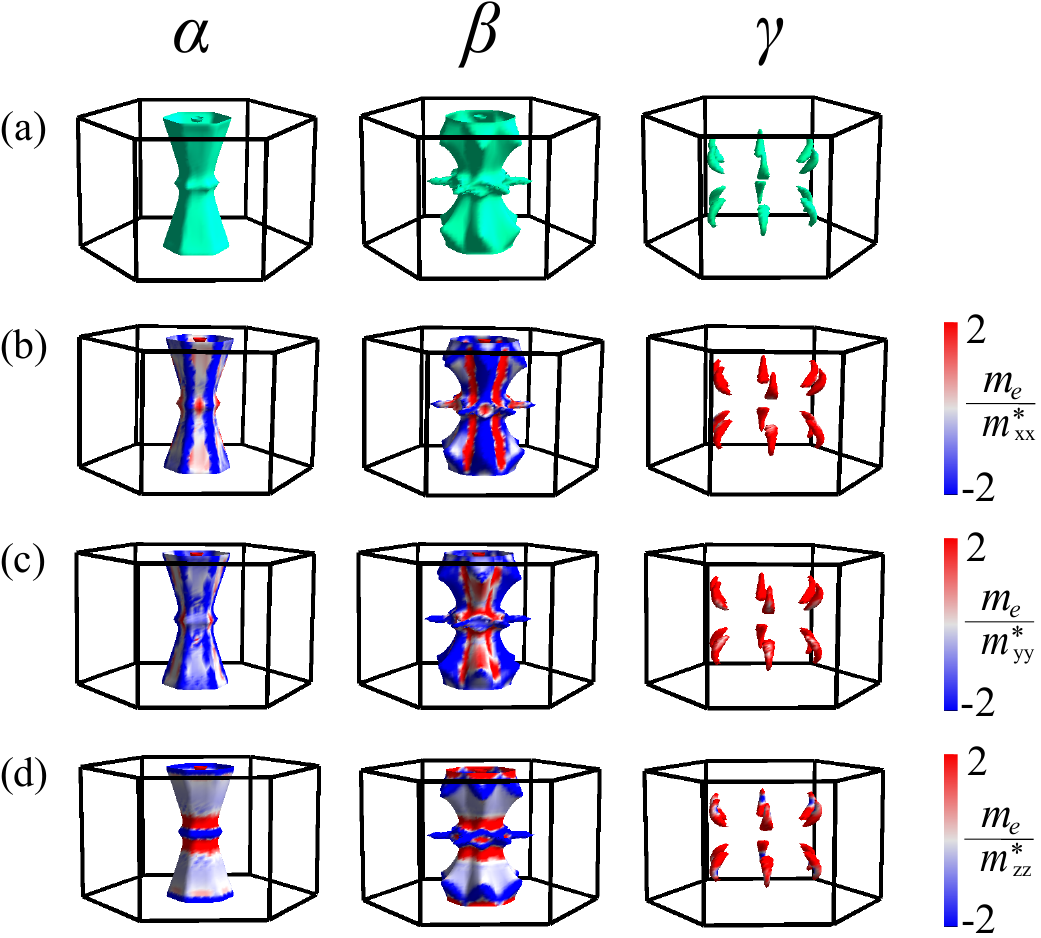}
\caption{(a) Calculated Fermi pockets $\alpha$, $\beta$, and $\gamma$ associated with three distinct bands $\#1$, $\#2$, and $\#3$, respectively, that cross $E'_\mathrm{F}$. (b)-(d) The distribution of IEM strength on the Fermi pockets along (b) x-, (c) y-, and (d) z-directions.}
\label{fig5}
\end{figure}

We investigated the microscopic origin of direction-dependent changes in the nature of charge carriers involved in the transport of CrSb within DFT calculations. Figure \ref{fig4}a shows the band structure of CrSb, plotted along high-symmetry paths of the BZ (see the inset of Figure \ref{fig4}a) with spin-orbit coupling (SOC) turned on. Since all these paths lie within the spin-degenerate nodal planes, which are protected by the sublattice-transposing symmetry of the CrSb crystal (see Supporting Information), the bands along these paths do not experience altermagnetic spin splitting in the absence of SOC.\cite{ding2024large,Zeng2024,Yang2024} However, when the SOC is included, the spin degeneracies in these nodal planes are lifted, leaving only a single nodal line along $\Gamma$–A. One finds a small spin polarization with in-plane spin components $\langle m_x \rangle$ and $\langle m_y \rangle$ along $\Gamma$–M, M–K, and K–$\Gamma$ directions (see Figure S7), a phenomenon referred to as weak altermagnetism.\cite{Yang2024,Krempask2024}
Real crystals often deviate from perfect stoichiometry, which changes the position of the Fermi energy. This must, however, be considered in the DFT calculations to capture the trends seen in the experiment. We find that adding a small number of holes per unit cell ($\sim0.13/\mathrm{u.c.}$) induces DDCP. In this case, the Fermi energy $E_\mathrm{F}$ is shifted below by 100 meV to $E'_\mathrm{F}$, represented by the black dashed line in Figure 4(a) and 4(b). To understand the unusual experimental results, we theoretically calculated the thermopower and the Hall coefficient tensor within the DFT framework using the Boltzmann transport equations and the relaxation-time approximation. The calculated Seebeck coefficient at 100 K is plotted as a function of the change in chemical potential in Figure \ref{fig4}b. It shows that the DDCP exists within a narrow energy window ($\sim14$ meV) around  $E'_\mathrm{F}$.  The FS obtained at $E'_\mathrm{F}$ (see Figure \ref{fig5}) is in good agreement with the FS observed in ARPES studies on CrSb,\cite{ding2024large,Zeng2024,Yang2024} confirming that Fermi energy is indeed shifted below its theoretical value in this system.

The temperature-dependent in-plane ($S_\mathrm{xx}$) and cross-plane ($S_\mathrm{zz}$) Seebeck coefficients are presented in Figure \ref{fig4}c. Within the examined temperature range, the absolute value of $S_\mathrm{xx}$ exhibits a slight, monotonic increase with increasing temperature. Conversely, the absolute value of $S_\mathrm{zz}$ initially rises with temperature, reaches a broad peak, and then decreases. However, $S_\mathrm{xx}$ remains negative across this temperature range, while $S_\mathrm{zz}$ is positive, which is qualitatively consistent with the experimental observations. We do not get quantitative agreement in the temperature dependence of the experimental data due to the approximations made while computing the Seebeck coefficients. The effects of phonons have been ignored in the description. \cite{Gupta2019,madsen2018boltztrap2} These could enter through electron-phonon interactions as well as through phonon drag effects.\cite{vaks1981singularities} Figure \ref{fig4}d shows the computed Hall coefficient components as a function of temperature. While $R_\mathrm{yxz}$ remains negative throughout the chosen temperature range, $R_\mathrm{xzy}$ and $R_\mathrm{zyx}$ stay positive. The constraint $R_\mathrm{xzy}=R_\mathrm{zyx}$, dictated by the magnetic point group symmetry of CrSb, is also accurately reflected. The trends in Hall coefficients are also consistent with the experimental observations.  These results suggest that the in-plane conduction is electron-dominated, while the cross-plane conduction is hole-dominated.

To understand the mechanism of DDCP in CrSb, we calculated the FS at $E'_\mathrm{F}$, as shown in Figure \ref{fig5}a. Three distinct Fermi pockets labeled $\alpha$, $\beta$, and $\gamma$ correspond to three different bands ($\#1$, $\#2$, and $\#3$, respectively) that cross $E'_\mathrm{F}$. Although band $\#3$ does not cross $E'_\mathrm{F}$ at $k_{z}=0 $ plane, it crosses $E'_\mathrm{F}$ at some finite value of $k_{z}$, as shown in Figure S8. The $\alpha$ and $\beta$ pockets appear to be nearly hyperboloid in shape, meaning they are closed in the $k_\mathrm{x}-k_\mathrm{y}$ plane and open along the $k_\mathrm{z}$ direction, apparently exhibiting both convex and concave features. Each of these pockets includes a small hemispherical sub-pocket around \textit{A} point of the BZ. The $\gamma$ pocket consists of 12 symmetry-related similar sub-pockets. These sub-pockets are closed and have an anisotropic, bead-like shape, elongated along the $k_\mathrm{z}$-direction.

In the presence of multiple Fermi pockets with
varying geometries, including hyperboloid-like and anisotropic features, the mechanism of DDCP in CrSb may not be determined simply by observing the shape of the FS. We therefore investigated the origin of this behavior by calculating the inverse effective mass (IEM) tensor, $m_{ij}^{*^{-1}}$, at each point of the FS. This involves the curvature of the electronic band dispersion at that point, as described by Eq. (1).
\begin{table}[b]
\centering
\caption{\label{tab1}
Average normalized inverse effective mass (IEM) values for the individual Fermi pockets $\alpha$, $\beta$, $\gamma$ and the entire FS along different directions.}
\begin{tabularx}{0.7\columnwidth}{c|X|X|X|X} 
  & $\alpha$ & $\beta$ & $\gamma$ & FS \\ \hline
$\left\langle \dfrac{m_e}{m_\mathrm{xx}^*}\right\rangle$ & -1.65 & -1.16 & 4.42 & 0.04 \\ 
$\left\langle \dfrac{m_e}{m_\mathrm{yy}^*}\right\rangle$ & -1.63 & -1.15 & 4.37 & 0.05 \\ 
$\left\langle \dfrac{m_e}{m_\mathrm{zz}^*}\right\rangle$ & -0.47 & -2.33 & 4.55 & -1.56 \\ \hline
\end{tabularx}
\end{table}

Figure \ref{fig5}b-d shows the distribution of the diagonal components of the IEM tensor, normalized by the free electron mass ($m_e$), on the three Fermi pockets. The pockets $\alpha$ and $\beta$ host both positive and negative distribution of IEM along both in-plane (x and y) and cross-plane (z) directions, with negative IEM apparently dominating the overall distribution. In contrast, the pocket $\gamma$ has predominantly positive IEM along all directions, indicating that it is an electron pocket. Notably, the small hemispherical $\alpha$ and $\beta$ sub-pockets also display positive IEM along all directions, indicating that they are electron pockets as well. Table \ref{tab1} presents the IEM values averaged over each of the three individual Fermi pockets and over the entire FS along x, y and z-directions. From this table, it is evident that both the $\alpha$ and $\beta$ pockets do not produce DDCP on their own (goniopolarity) as they have an overall hole character (negative effective mass) along all the directions. The average IEM values for the $\gamma$ pocket along all directions are positive which means it has overall electron character along all the directions. When the entire FS is considered, the in-plane average IEM values become positive, while the cross-plane average IEM value becomes negative, giving rise to DDCP. This suggests that DDCP in CrSb arises from the multicarrier mechanism, where in-plane conduction is dominated by electrons and cross-plane conduction is dominated by holes.

\begin{figure}[t]
\centering
\includegraphics[width=0.7\linewidth]{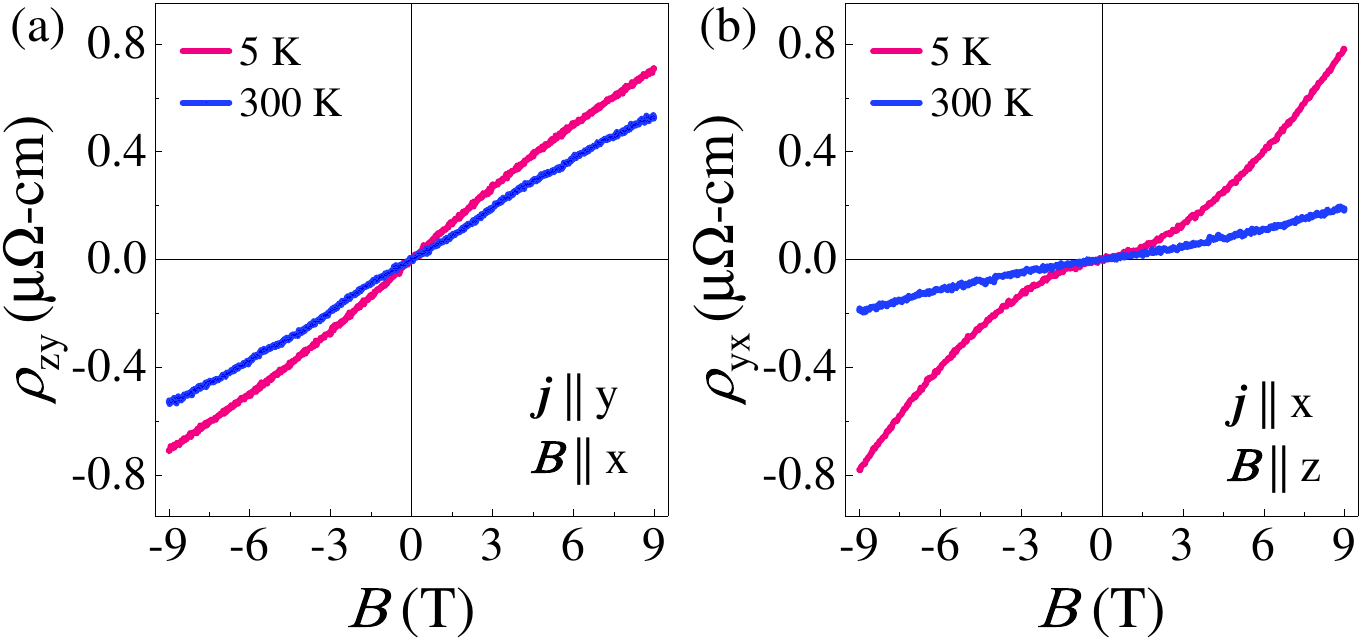}
\caption{\label{fig7} Magnetic field dependent Hall resistivity of Cr$_{0.98}$V$_{0.02}$Sb at temperatures 5 K and 300 K for (a) in plane ($B||x$) and (b) out-of-plane ($B||z$) magnetic fields. }
\end{figure}

Our DFT calculations predict that DDCP exists only within a small energy range ($\Delta E\approx 14\:\mathrm{meV}$) around $E'_\mathrm{F}$. To test the validity of this, we doped CrSb with $2\%$ V and synthesized   Cr$_{0.98}$V$_{0.02}$Sb. This introduces an additional 0.04 number of holes per unit cell which is expected to move the system out of the DDCP window. Figure \ref{fig7}a and \ref{fig7}b show the field-dependent Hall resistivity $\rho_\mathrm{zy}$ and $\rho_\mathrm{yx}$, respectively for Cr$_{0.98}$V$_{0.02}$Sb at temperatures 5 K and 300 K. While the hole doping has little effect on $\rho_\mathrm{zy}$, as evidenced by its similar positive slopes for CrSb and Cr$_{0.98}$V$_{0.02}$Sb (see 300 K data), it significantly modifies $\rho_\mathrm{yx}$, which retains non-linearity but now exhibits positive slopes, indicating hole-dominant conduction. A consistent result is also obtained in the thermopower measurement (see Figure S10) where the in-plane Seebeck coefficient, $S_\mathrm{xx}$, has changed sign and become positive for Cr$_{0.98}$V$_{0.02}$Sb. This confirms that even a small amount of hole doping suppresses the DDCP and renders the system p-type, consistent with our DFT predictions.

\section{Conclusion}
In summary, we have studied the anisotropic transport properties of altermagnetic CrSb. Both Hall and Seebeck coefficients indicate that CrSb exhibits DDCP, with in-plane conduction dominated by electrons and cross-plane conduction dominated by holes. Our DFT calculations based on the Boltzmann transport equation and the relaxation-time approximation qualitatively support our experimental results. We show that the DDCP in CrSb arises from the multicarrier mechanism and identify two hole pockets and one electron pocket dominating the cross-plane and in-plane conduction, respectively. The emergence of DDCP in CrSb is a consequence of its specific Fermi surface geometry, which is shaped by altermagnetic spin splitting. However, it is important to emphasize that altermagnetic spin splitting does not inherently lead to DDCP in all materials. In CrSb, the particular Fermi surface geometry happens to facilitate DDCP, but this behavior is not a universal feature of altermagnets. Furthermore, our calculations predict that the DDCP in CrSb is fragile and can be suppressed with a small amount of doping. This prediction is validated by the loss of DDCP in minimally hole doped Cr$_{0.98}$V$_{0.02}$Sb. Composed of earth-abundant and non-toxic elements, CrSb emerges as a promising candidate for practical applications leveraging the DDCP behavior. Furthermore, the inherent altermagnetic nature of CrSb, characterized by large spin-splitting, opens up exciting possibilities for future spintronic applications that can exploit both DDCP and spin-splitting phenomena.

\section{Method}
\textit{Crystal growth and characterization}: Single crystals of CrSb and Cr$_{0.98}$V$_{0.02}$Sb were grown by the chemical vapor transport (CVT) method. To grow CrSb, Cr (99.99\%, \textit{Alfa Aesar}) and Sb powders (99.8\%, \textit{Alfa Aesar}) (1 g in total) were taken in a $1:1$ molar ratio together with 100 mg of iodine and sealed in a quartz tube 160 mm long and 15 mm in diameter. The tube was then evacuated, sealed under a partial Ar atmosphere, and placed in a horizontal three-zone tube furnace for 10 days. A uniform temperature gradient was maintained along the tube for the transport process, keeping the reactant end of the tube at 750 \textdegree C and the product end at 650 \textdegree C.  Large hexagonal plate-like single crystals up to 1 mm thick and 5 mm wide were formed near the product end of the tube at the end of the reaction. Single crystals of Cr$_{0.98}$V$_{0.02}$Sb were grown following the same process with the starting ratio of elements. 

The elemental composition was verified using energy dispersive x-ray spectroscopy (EDXS) (see Supporting Information). The EDXS data were collected on a field emission electron microscope (Quanta 250 FEG) equipped with an Element silicon drift detector (SDD) with an accelerating voltage of 25 kV and an accumulation time of 60 seconds. X-ray diffraction (XRD) experiments were performed using a Rigaku SmartLab diffractometer equipped with a 9 kW Cu $k_\alpha$ X-ray source. The phase purity of CrSb was verified using powder X-ray diffraction, where the pattern was fitted using the Le-Bail method (see Supporting Information).

\textit{Electrical transport and thermopower experiments}: The magnetotransport measurements were carried out using the ETO (electrical transport option) option of the Physical Properties Measurement System (PPMS, 9\,T, Dynacool, Quantum Design). A standard six-terminal method was employed for simultaneous measurements of longitudinal resistivity $(\rho_{ii})$ and transverse Hall resistivity $(\rho_{ij})$. The indices $i$ and $j$ in $\rho_{ij}$ are the direction of measured voltage and applied current, respectively. Proper symmetrisation and anti-symmetrisation were done to get rid of any additional contribution coming from possible misalignment of the contact probes.

The thermopower measurements were carried out on a bulk single crystal in both directions. The two probe electrodes were made on top of the crystal sample by fixing Cu-wires using silver paste. These measurements were performed in a liquid nitrogen cryostat where the sample was kept under vacuum. The temperature difference between two ends of the sample was generated by sending currents from a source meter (Keithley 2450) to the heater attached just beside the sample. The voltage and temperature difference between two ends of the sample were measured by a precision nano voltmeter (Keithley 2182A) and temperature controller (Lakeshore 340). The bath temperature is controlled using a temperature controller (Lakeshore 340) with temperature fluctuations less than $\pm20\,\mathrm{mK}$.

\textit{DFT Calculations}: The electronic structure was computed within the Vienna Ab initio Simulation Package (VASP), employing a plane-wave basis set and projected augmented wave potentials.\cite{blochl1994projector,kresse1993ab,kresse1994ab,kresse1996efficient} 
The generalized gradient approximation (GGA) \cite{perdew1996generalized} for the exchange-correlation functional was used. While the lattice parameters were fixed at the experimental values \cite{KALLEL19741139} for the NiAs structure of CrSb in the $P6_3/mmc$ space group,  all internal atomic positions were optimized through a total energy minimization. A cutoff energy of 500 eV was used to define the maximum kinetic energy for the plane waves in the basis. Brillouin zone integrations were performed with a Monkhorst-Pack \textit{k}-grid of density $10 \times 10 \times 10$  points.\cite{monkhorst1976special} The Seebeck and Hall coefficients were calculated using the semiclassical Boltzmann transport equation. These calculations employed the relaxation time and rigid band approximations, with the DFT band dispersions interpolated on a dense 35×35×35 \textit{k}-mesh grid as implemented in Boltztrap2.\cite{madsen2018boltztrap2} Additionally, the inverse effective mass (IEM) tensor was determined using B-spline interpolation of the DFT band dispersions.


\section*{Acknowledgements}
This work utilized the instrumentation facilities provided by the Technical Research Centre (TRC) at S. N. Bose National Centre for Basic Sciences, under Department of Science and Technology (DST), Government of India. NK acknowledges the Science and Engineering Research Board (SERB), India, for financial support through Grant Sanction No. CRG/2021/002747 and Max Planck Society for funding under Max Planck-India partner group project. PM acknowledges support from SERB through the project SERB-POWER (SPF/2021/000066). MM acknowledges CSIR-Human Resource Development Group (HRDG) (03/1511/23/EMR-II). BR acknowledges the financial support from DST. KP acknowledges Param Rudra Facility under the National Supercomputing Mission, Government of India at S. N. Bose National Centre for Basic Sciences.

\section*{Conflict of interest}
The authors declare no conflict of interset.


\input{main_text.bbl}
\newpage
\input{Supporting_information}


\end{document}

%% file: Supporting_information.tex

\captionsetup[figure]{labelformat=simple, labelsep=period, name=Figure}

\captionsetup[figure]{labelformat=simple, labelsep=period, name=Figure}
\begin{center}
    \textbf{\Large Supporting Information}\\
    \vspace{1cm}
     \textbf{\Large Direction-Dependent Conduction Polarity in Altermagnetic CrSb}\\
     \vspace{5mm}
     Banik Rai\textsuperscript{1,3}, Krishnendu Patra\textsuperscript{1,3}, Satyabrata Bera\textsuperscript{2}, Sk Kalimuddin\textsuperscript{2}, Kakan Deb\textsuperscript{1},\\ Mintu Mondal\textsuperscript{2}, Priya Mahadevan\textsuperscript{1}, Nitesh Kumar\textsuperscript{1}\\
     \vspace{2mm}
     \textsuperscript{1}Department of Condensed Matter and Materials Physics, S. N. Bose National Centre for Basic Sciences, Salt Lake City, Kolkata-700106, India\\
      \textsuperscript{2}School of Physical Sciences, Indian Association for the Cultivation of Science, Jadavpur, Kolkata-700032, India\\
      \textsuperscript{3}These authors contributed equally.
\end{center}
\vspace{1cm}

\section*{S1 \hspace{5mm}EDXS and XRD}
Figure S1 shows the typical EDXS spectrum of the elemental composition of CrSb and Cr$_{0.98}$V$_{0.02}$Sb. The EDXS results, obtained from several spots on single crystals of CrSb and Cr$_{0.98}$V$_{0.02}$Sb, are tabulated in Table S1. The elemental ratio for CrSb is close to 1:1.  Upon doping with V, the nominal EDXS composition is nearly Cr$_{0.98}$V$_{0.02}$Sb.

\begin{figure}[h]
\renewcommand{\thefigure}{S1}
\centering
\includegraphics[width=0.55\columnwidth]{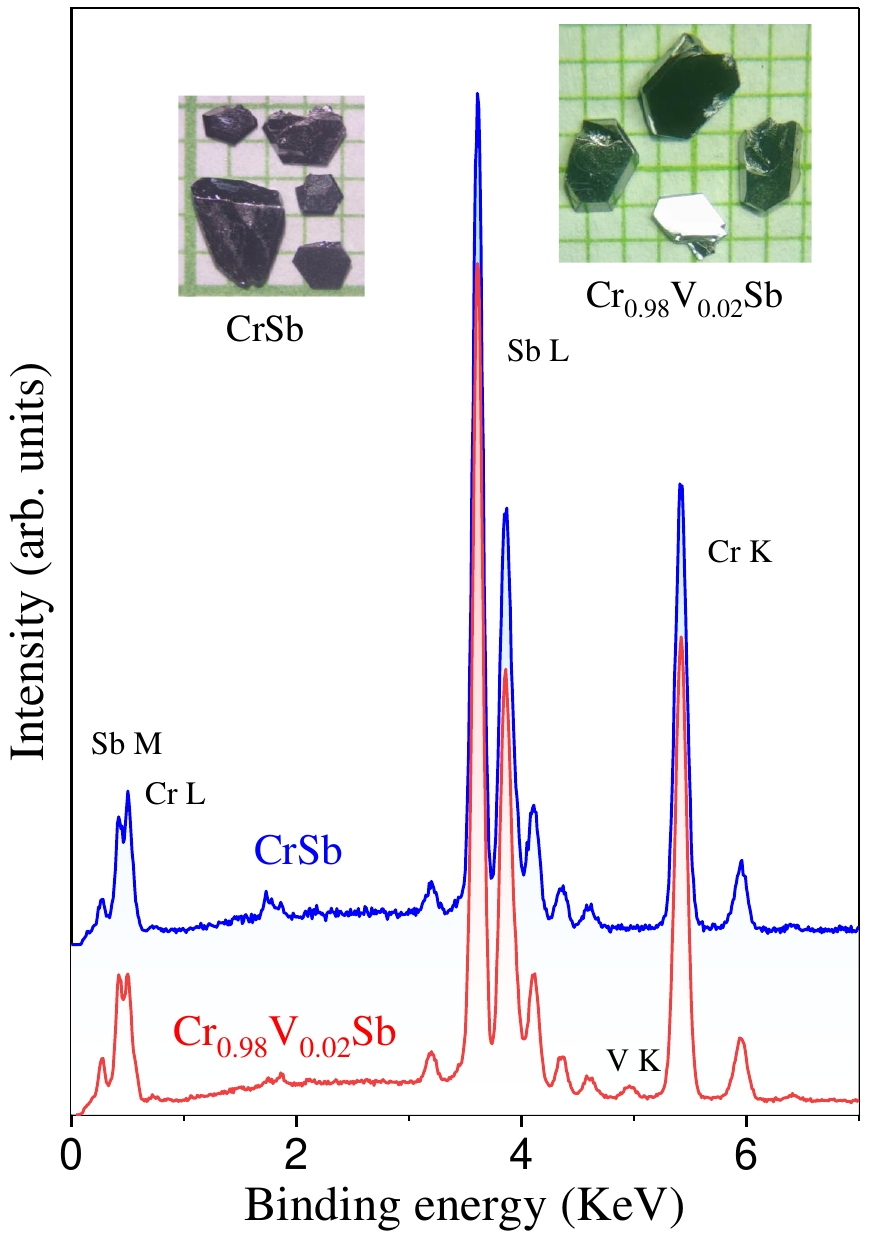}
\caption{EDXS spectrum of CrSb (blue) and Cr$_{0.98}$V$_{0.02}$Sb (red).}
\label{Fig s1}
\end{figure}

\begin{table}[h!]
\centering
\renewcommand{\thetable}{S1} 
\caption{Chemical composition and derived chemical formulas.}
\begin{tabular}{cccccccc}
\hline
Compound & \%Cr & \%V & \%Sb & \(\langle \%Cr \rangle\) & \(\langle \%V \rangle\) & \(\langle \%Sb \rangle\) & Chemical Formula \\ 
\hline
\multirow{4}{*}{CrSb} 
& 49.1 & - & 50.1 & \multirow{4}{*}{50.325} & \multirow{4}{*}{-} & \multirow{4}{*}{49.675} & \multirow{4}{*}{Cr$_{1.0065}$Sb$_{0.9935}$} \\
& 50.3 & - & 49.7 & & & & \\
& 51.0 & - & 49.0 & & & & \\
& 50.9 & - & 49.1 & & & & \\
\hline
\multirow{5}{*}{Cr$_{0.98}$V$_{0.02}$Sb} 
& 49.5 & 0.8 & 49.7 & \multirow{5}{*}{49.26} & \multirow{5}{*}{0.84} & \multirow{5}{*}{49.86} & \multirow{5}{*}{Cr$_{0.985(6)}$V$_{0.017(5)}$Sb$_{0.997(2)}$} \\
& 49.1 & 1.0 & 49.9 & & & & \\
& 49.6 & 0.5 & 49.9 & & & & \\
& 48.8 & 1.2 & 49.9 & & & & \\
& 49.3 & 0.7 & 49.9 & & & & \\
\hline
\end{tabular}
\end{table}

Figure S2(a) shows the powder XRD pattern of CrSb fitted with the space group $P63/mmc$ using the Le-bail method. Fig. S2(b) shows the XRD pattern of a plate-like single crystal of CrSb. The observation of only $(000l)$ peaks confirms that the surface exposed to the XRD beam is \textit{ab}-plane and the absence of other peaks indicates good crystallinity of the single crystal.
\begin{figure}[h]
\renewcommand{\thefigure}{S2}
\centering
\vspace{0.4 em}
\includegraphics[width=0.8\columnwidth]{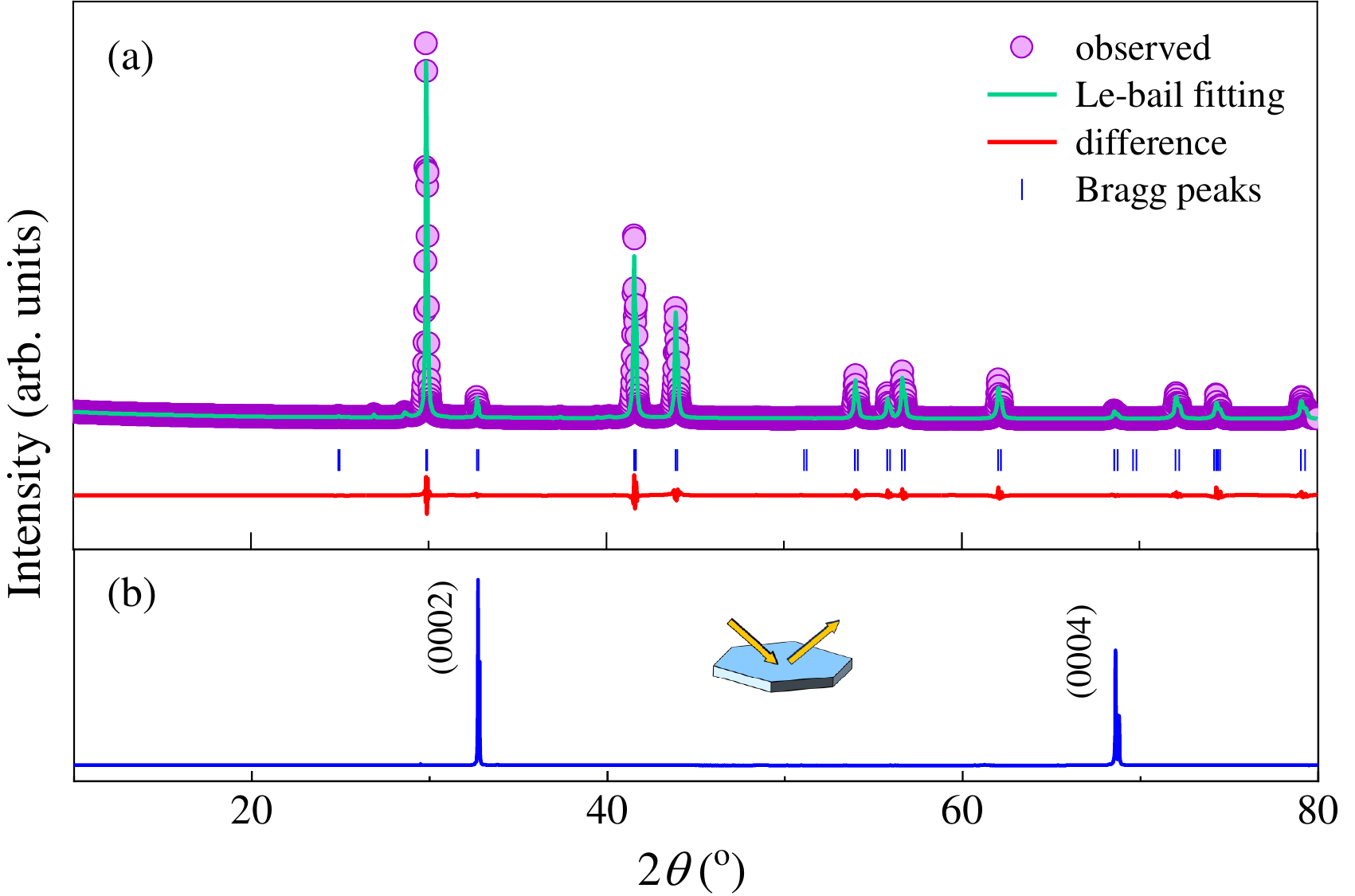}
\caption{XRD study on CrSb. (a) Powder XRD pattern of CrSb fitted with space group $P6_3/mmc$ using the Le-bail method. (b) XRD pattern of a plate-like single crystal of CrSb.}
\label{Fig s2}
\end{figure}
\section*{S2 \hspace{5mm}Magnetoresistance and multi-carrier fitting}
Figure S3(a)-(c) shows the magnetic field-dependent transverse magnetoresistance $\bigg[\mathrm{MR}=\dfrac{\rho_{ii}(B)-\rho_{ii}(0)}{\rho_{ii}(0)}\times 100\:\%\bigg]$ of CrSb at different temperatures, measured along the three orthogonal x, y, and z directions. MR reaches its maximum value of 8\% for $B||\mathrm{z}$ [Figure S3(c)] at 9 T and 2 K. For the other two directions, the MR is smaller, with the smallest value observed for $B||\mathrm{x}$ [Fig. S3(a)]. In all directions, the observed positive MR can be attributed to the orbital cyclotron motion of charge carriers under the applied magnetic field. In systems with nearly equal electron and hole carrier densities (charge compensation), the MR typically exhibits a quadratic dependence on the magnetic field.\cite{ashcroft1978solid} Figure S3(d), showing the MR at 2 K for all three directions, highlights the deviation of the MR from the quadratic behavior, indicating the presence of uncompensated charge carriers in CrSb.

In materials with multiple bands crossing the Fermi level ($E_\mathrm{F}$), the mobility ($\mu$) of charge carriers often varies between the bands. As a result, the charge carriers move independently within these bands, each characterized by its own carrier density ($n_i$) and mobility ($\mu_i$).\cite{hurd2012hall} This results in a non-linear Hall signal as a function of the magnetic field. The longitudinal ($\sigma_{\mathrm{xx}}$) and Hall ($\sigma_{\mathrm{xy}}$) conductivity, considering the multi-carrier effect, can be expressed as:

\begin{figure}[t]
\renewcommand{\thefigure}{S3}
\centering
\includegraphics[width=0.7\columnwidth]{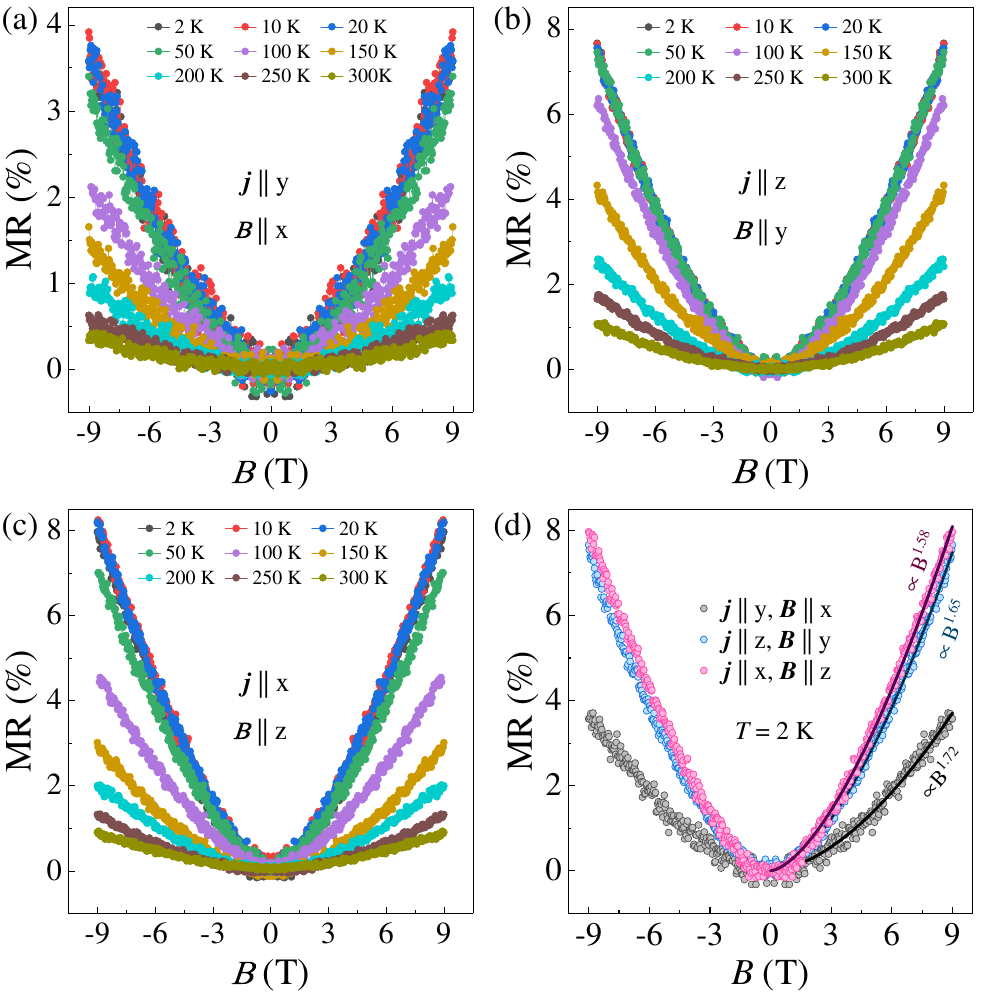}
\caption{(a)-(c) Magnetoresistance of CrSb measured in various geometries at different temperatures. (d) Magnetoresistance at 2K measured along various directions and the corresponding exponential fit.}
\label{Fig s2}
\end{figure}

\begin{align}
    \sigma_{\mathrm{xx}}&=e \sum \frac{n_i\mu_i}{1+\mu_i^2B^2} \tag{S1}\\
    \sigma_{\mathrm{xy}}&=eB \sum \frac{n_i\mu_i^2}{1+\mu_i^2B^2} \tag{S2} .
\end{align}

\begin{table}[b]
\centering
\renewcommand{\thetable}{S2} 
\caption{Carrier density and mobility at various temperatures obtained by three carrier fit.}
\begin{tabular}{c|c|c|c|c|c|c}
Temperature & $n_\mathrm{h}$ & ${n}_\mathrm{e1}$ & ${n}_\mathrm{e2}$ & $\mu_\mathrm{h}$ & $\mu_\mathrm{e1}$ & $\mu_\mathrm{e2}$\\
(K) & $(10^{21}\mathrm{cm}^{-3})$ & $(10^{20}\mathrm{cm}^{-3})$ & $(10^{19}\mathrm{cm}^{-3})$ & $(\mathrm{cm}^2\mathrm{V}^{-1}\mathrm{s}^{-1})$ & $(\mathrm{cm}^2\mathrm{V}^{-1}\mathrm{s}^{-1})$ & $(\mathrm{cm}^2\mathrm{V}^{-1}\mathrm{s}^{-1})$ \\\hline
2  & 1.45  & 6.39 & 1.75 & 312 & 480 & 2394\\ 
10  & 1.42  & 6.61 & 1.84 & 316 & 471 & 2344\\ 
20  & 1.39  & 6.78 & 1.87 & 315 & 458 & 2304\\ 
50  & 1.44  & 6.20 & 1.30 & 264 & 435 & 2196\\ 
100 & 1.12  & 7.63 & 0.89 & 204 & 281 & 1674\\
150 & 0.69  & 11.2 & 0.69 & 186 & 165 & 1293\\
\hline
\end{tabular}
\end{table}

Figure S4(a)-(b) shows the field-dependent $\sigma_{\mathrm{xx}}$ and $\sigma_{\mathrm{xy}}$ at 2 K fitted using equation (S1) and (S2), considering two and three carrier models. The fitting quality for two-carrier model is poor, indicating the presence of more than two independent types of charge carriers. The fitting improves significantly with the three-carrier model. This model also fits the data at higher temperatures, as shown in Figure S4(c)-(d). The values of $n$ and $\mu$ obtained from the fit are presented in Figures S4(e) and S4(f), respectively. Of the three independent charge carriers, one carrier type is hole-like (\textit{h}) and the other two are electron-like (\textit{e}1 and \textit{e}2). The hole-like carrier has the highest carrier density ($n_{h}\approx 1\times10^{21}\mathrm{cm}^{-3}$). The two electron-like carriers have smaller carrier density, with one being much smaller than the other. The mobilities follow the opposite trend. The hole mobility $(\mu_{h})$ is the smallest and the mobility of the second electron-like carrier ($\mu_{e2}$) is the largest, reaching a value as large as 2394 $\mathrm{cm^2V^{-1}s^{-1}}$ at 2 K. Table S2 shows the variation of carrier density and mobility of three carriers with temperature as obtained from the three-carrier fit. 

\begin{figure}[t]
\renewcommand{\thefigure}{S4}
\centering
\includegraphics[width=1\columnwidth]{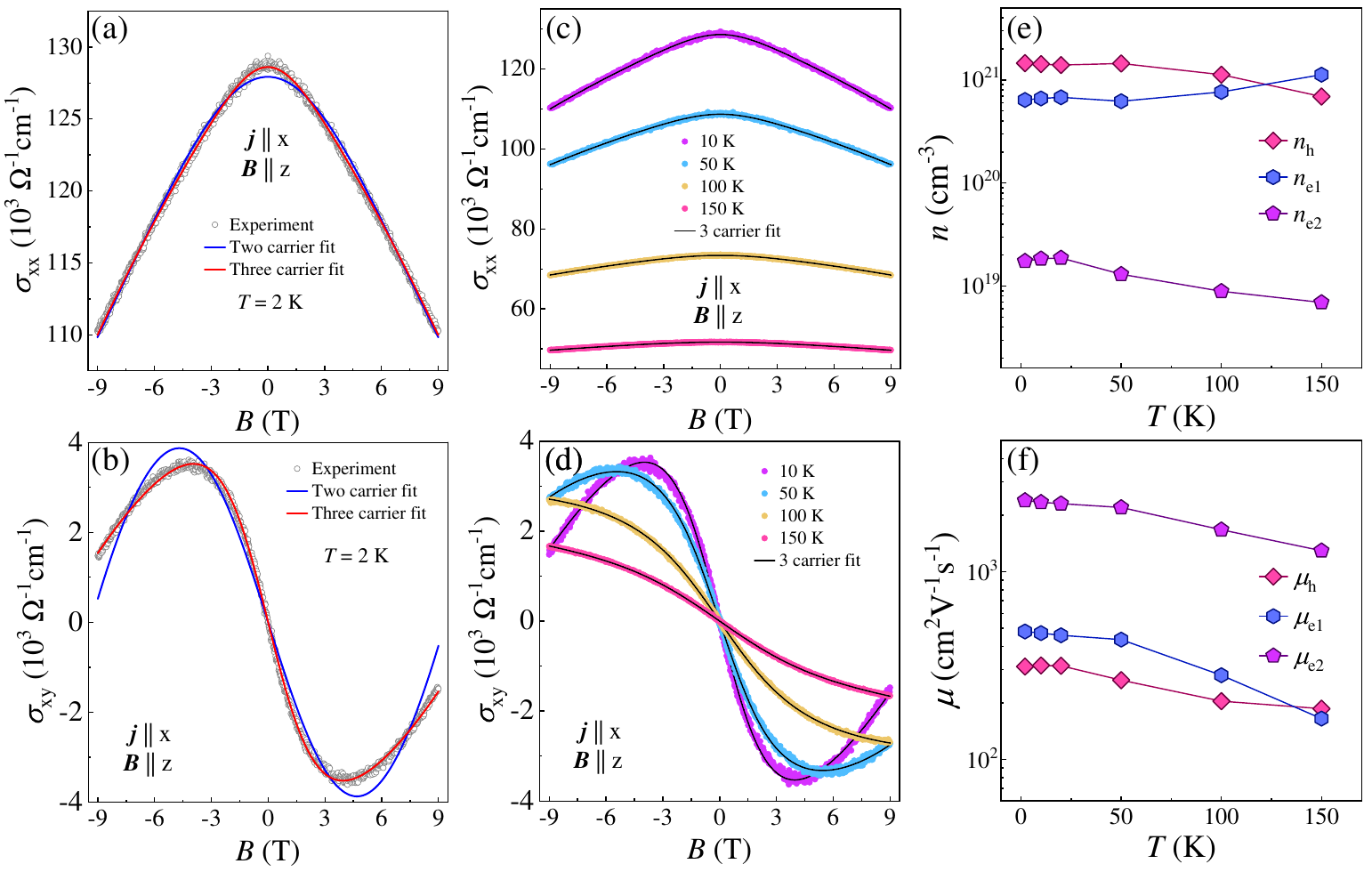}
\caption{Multicarrier effect in the transport of CrSb. (a)-(b) Global (simultaneous) fitting of longitudinal ($\sigma_{\mathrm{xx}}$) and Hall ($\sigma_{\mathrm{xy}}$) conductivity at 2 K using two-carrier and three-carrier model. (c)-(d) Global fitting of $\sigma_{\mathrm{xx}}$ and $\sigma_{\mathrm{xy}}$ at various temperatures using three-carrier model. Variation of (e) carrier density and (f) mobility with temperature as obtained from three carrier fit. These values are in close agreement with the values reported in Ref. \cite{bai2024nonlinear}.}
\label{Fig s1}
\end{figure}
\begin{figure}[!b]
\renewcommand{\thefigure}{S5}
\centering
\includegraphics[width=0.8\columnwidth]{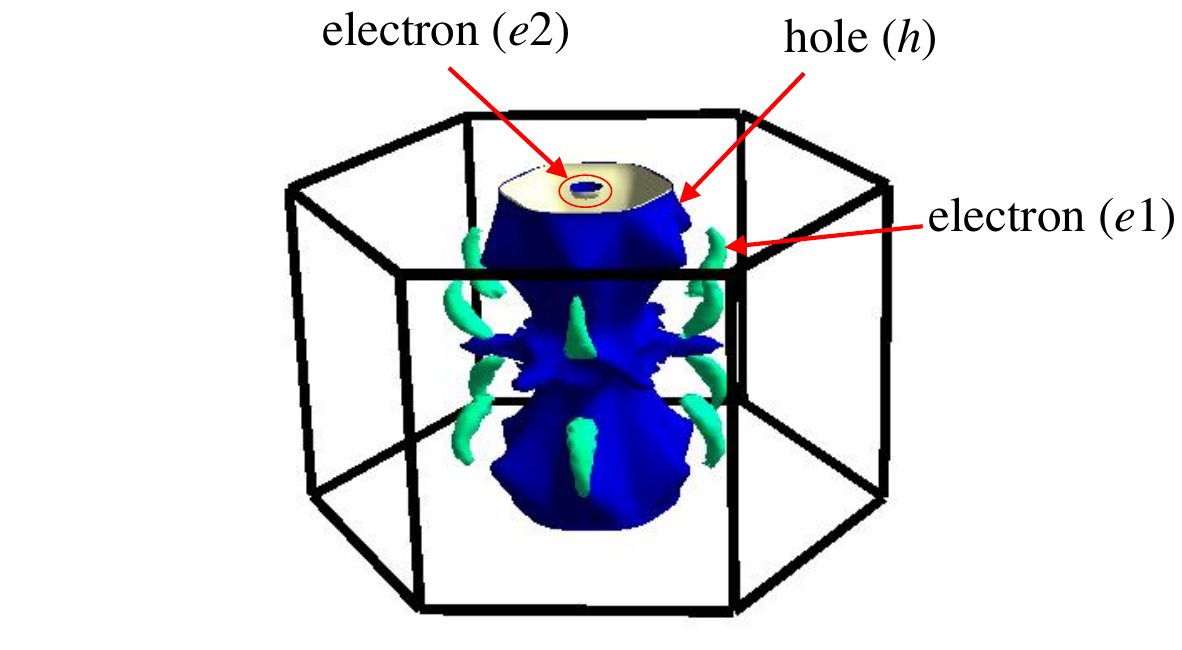}
\caption{The attribution of different carrier types, as obtained from the three-carrier fit, to different portions of the FS.}
\label{Fig s1}
\end{figure}
The Fermi surface (FS) of CrSb consists of three pockets, namely $\alpha$, $\beta$, and $\gamma$, corresponding to three distinct bands crossing $E_\mathrm{F}$ (see main text). Effective mass calculations (main text Table 1) reveal that the $\alpha$ and $\beta$ are hole pockets, while $\gamma$ is an electron pocket. This seemingly contradicts the results of the three-carrier fitting, which identifies two electron-like and one hole-like carriers. However, this apparent discrepancy can be resolved by considering the origin of the $\alpha$ and $\beta$ pockets. These pockets arise from the altermagnetic and spin-orbit splitting of the same energy band. Consequently, they exhibit strong similarities in their electronic properties. Furthermore, the effective masses of the charge carriers residing within these pockets are nearly identical, suggesting that their mobilities are also highly comparable. Due to these strong similarities, the charge carriers within the $\alpha$ and $\beta$ pockets do not behave as entirely independent entities in transport. Their contributions tend to be highly correlated, effectively merging into a single, dominant hole-like carrier contribution in the three-carrier fit. The first electron-like carrier (\textit{e}1) in the three-carrier fit can be attributed to the electron pocket $\gamma$. The second electron-like carrier (\textit{e}2) most likely correspond to the small hemispherical sub-pockets present within the $\alpha$ and $\beta$ pockets, as these sub-pockets are electron-like in nature (see main text). Since, these sub-pockets are well separated from the larger portion of the $\alpha$ (or $\beta$) pocket and exhibit distinct geometry, the charge carriers residing on them can contribute independently to the transport, even though they arise from the same energy band. The value of ${n_{e2}}$ is about 100 times smaller than ${n_{h}}$, as evident from Table S2. A similar proportion is visibly evident in the size of the larger portion of the $\alpha$ (or $\beta$) pocket and its sub-pocket. Since, carrier density is proportional to the volume of the corresponding Fermi pocket, our attribution of second electron-like carrier to the hemispherical sub-pockets is well supported. The value of $n_{e1}$ is also consistent with the relative size of the $\gamma$ pocket. In Figure S4, different carrier types, as obtained from the three-carrier fit, have been attributed to specific portions of the FS.

\section*{S3\hspace{5mm}Altermagnetic splitting of energy bands}
Altermagnets have been identified as a new phase of collinear magnets that share properties with both ferromagnets (e.g., non-relativistic spin splitting of energy bands) and antiferromagnets (e.g., zero net magnetization)\cite{smejkal2022}. In altermagnets, the spatial arrangement of non-magnetic atoms surrounding the magnetic ones plays a crucial role in dictating their distinct spin group symmetry, which distinguishes them from ferromagnets and conventional antiferromagnets. Unlike conventional antiferromagnets, where the opposite spin sublattices are connected by a simple lattice translation or inversion, in altermagnets, this connection is established by a real-space crystal rotation. The symmetries of an antiferromagnetic crystal, i.e.\ $\mathcal{T}$, ${\textit{\textbf{t}}}\mathcal{T}$ and $\mathcal{P}\mathcal{T}$, where $\mathcal{T}$, ${\textit{\textbf{t}}}$ and $\mathcal{P}$ are time reversal, lattice translation and space inversion operations, are thus broken in an altermagnet which facilitates the momentum dependent spin splitting of energy bands even without spin-orbit coupling akin to ferromagnets. 
For CrSb, the sublattice transposing symmetries are $C_\mathrm{6z}t_{1/2}\tau$ and $M_\mathrm{z}\tau$ (see main text). The first symmetry, $C_\mathrm{6z}t_{1/2}\tau$, results in three spin-degenerate nodal planes in the Brillouin zone (BZ) as shown by the blue vertical planes in Figure S6(a). The second symmetry, $M_\mathrm{z}\tau$, result in one spin-degenerate nodal plane in the BZ, as shown by a orange horizontal plane in Figure S6(a). Spin-degeneracy is lifted when moving away from these four nodal planes. Figure S6(b) shows the band structure of CrSb calculated along the path $\mathrm{L_1}-\Gamma-\mathrm{L_2}$ of the BZ in the absence of spin-orbit coupling (SOC). As this path lies outside the four nodal-planes, the bands experience significant splitting.

\begin{figure}[!b]
\renewcommand{\thefigure}{S6}
\centering
\includegraphics[width=0.8\columnwidth]{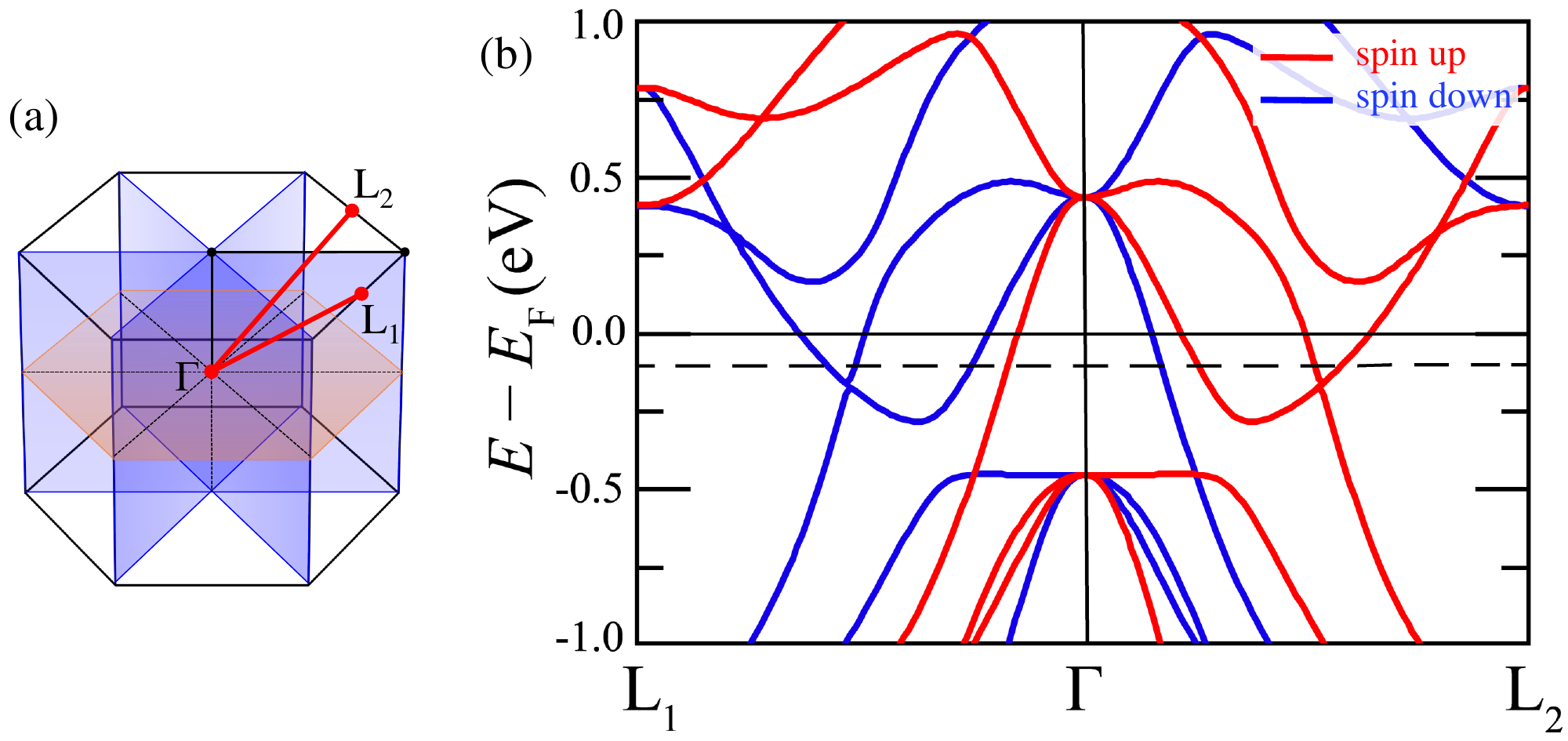}
\caption{(a) Brillouin zone (BZ) of CrSb showing four spin-degenerate nodal planes. (b) Band structure of CrSb calculated in the absence of spin-orbit coupling (SOC) along the paths as shown in Figure S6(a). A significant altermagnetic splitting can be observed. The dashed horizontal line represents the experimental fermi energy ($E'_\mathrm{F}$) (see main text).}
\label{Fig s5}
\end{figure}
\newpage

\section*{S4 \hspace{5mm}Magnetic component-projected relativistic band dispersion}

\FloatBarrier

\begin{figure}[H]
\renewcommand{\thefigure}{S7}
\centering
\includegraphics[width=0.98\columnwidth]{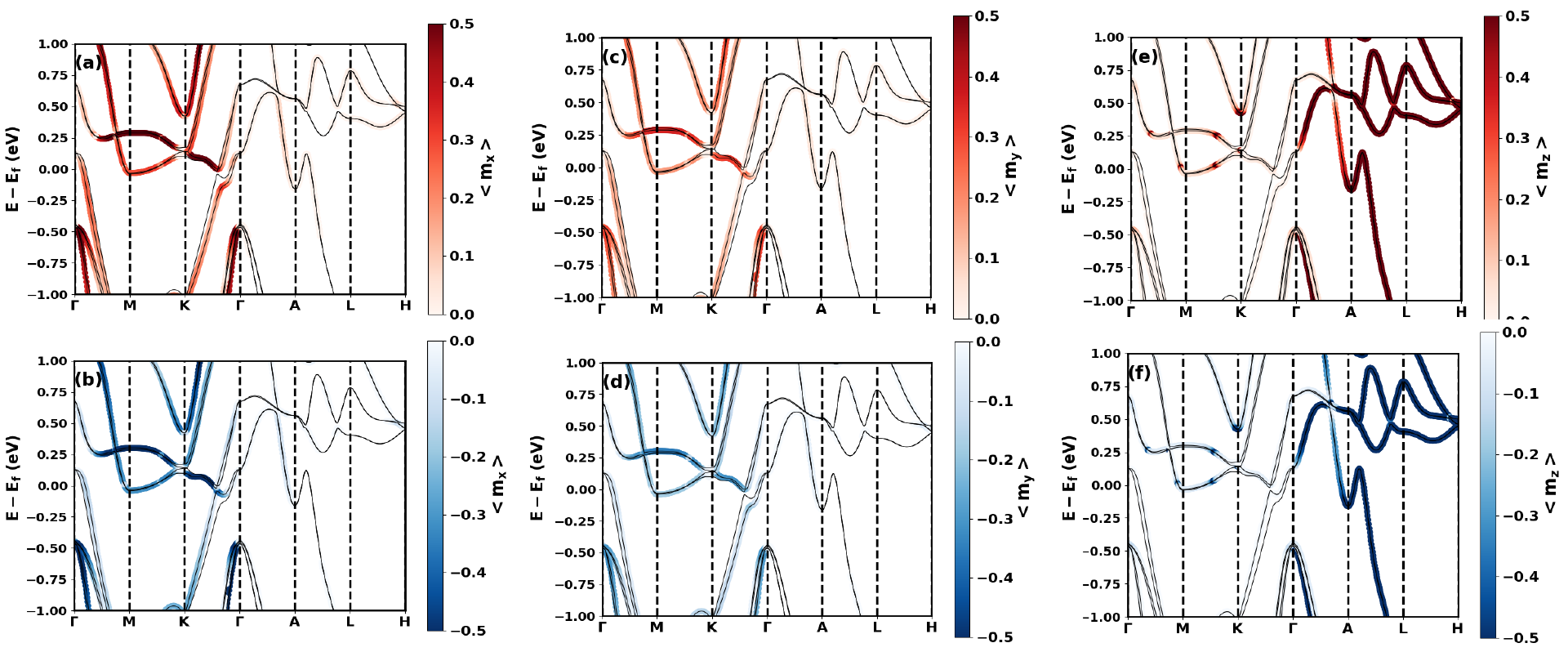}
\caption{Band dispersions for CrSb plotted along various symmetry directions. Spin-orbit interactions have been included in the calculation of the band structure. The spin projections for positive (a) $\langle\mathrm{m}_x\rangle$ (c) $\langle\mathrm{m}_y\rangle$ and (e) $\langle\mathrm{m}_z\rangle$ as well as negative (b) $\langle\mathrm{m}_x\rangle$ (d) $\langle\mathrm{m}_y\rangle$ and (f) $\langle\mathrm{m}_z\rangle$ have been shown. The color bar on the right of each plot gives the magnitude.
}
\label{Fig s6}
\end{figure}

\section*{S5 \hspace{5mm}Band dispersion at finite $k_{z}$ plane}
 It is evident from the FS that there is no $\gamma$ pocket at the $k_z = 0$ plane, as there is no intersection between band \#3 and the shifted Fermi energy ($E'_\mathrm{F}$) (indicated by dashed horizontal line) in that plane. However, at a finite $k_z$, band \#3 intersects $E'_\mathrm{F}$ which gives rise to the $\gamma$ pocket in the FS. In Figure S7, relativistic band dispersion is shown for the $k_z=0.2\pi/c$ plane, where the band \#3 intersects $E'_\mathrm{F}$.
 
\begin{figure}[h]
\renewcommand{\thefigure}{S8}
\centering
\includegraphics[width=0.65\columnwidth]
{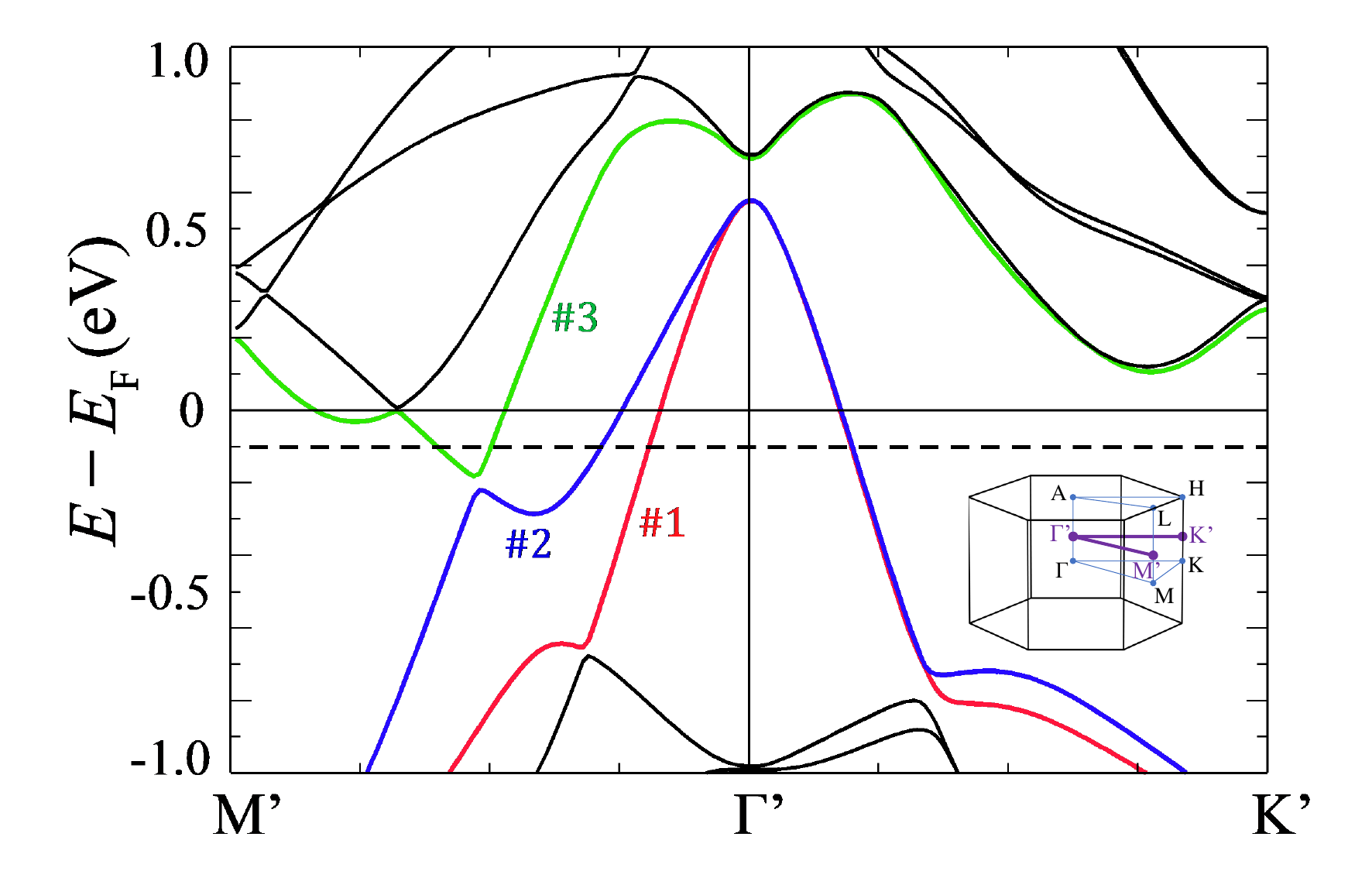}
\caption{Band dispersion (with SOC turned on) along high symmetry lines for a particular fixed plane $k_{z}=0.2\pi/c$.  }
\end{figure}

\section*{S6 \hspace{5mm}Number of electrons per unit cell (u.c.) as a function of energy.}

We have calculated the number of electrons at a particular energy using DFT calculations. From this, we determined the change in the number of electrons at the shifted Fermi energy ($E'_\mathrm{F}$) as shown in Figure S9. 
\begin{figure}[h]
\renewcommand{\thefigure}{S9}
\centering
\includegraphics[width=0.65\columnwidth]{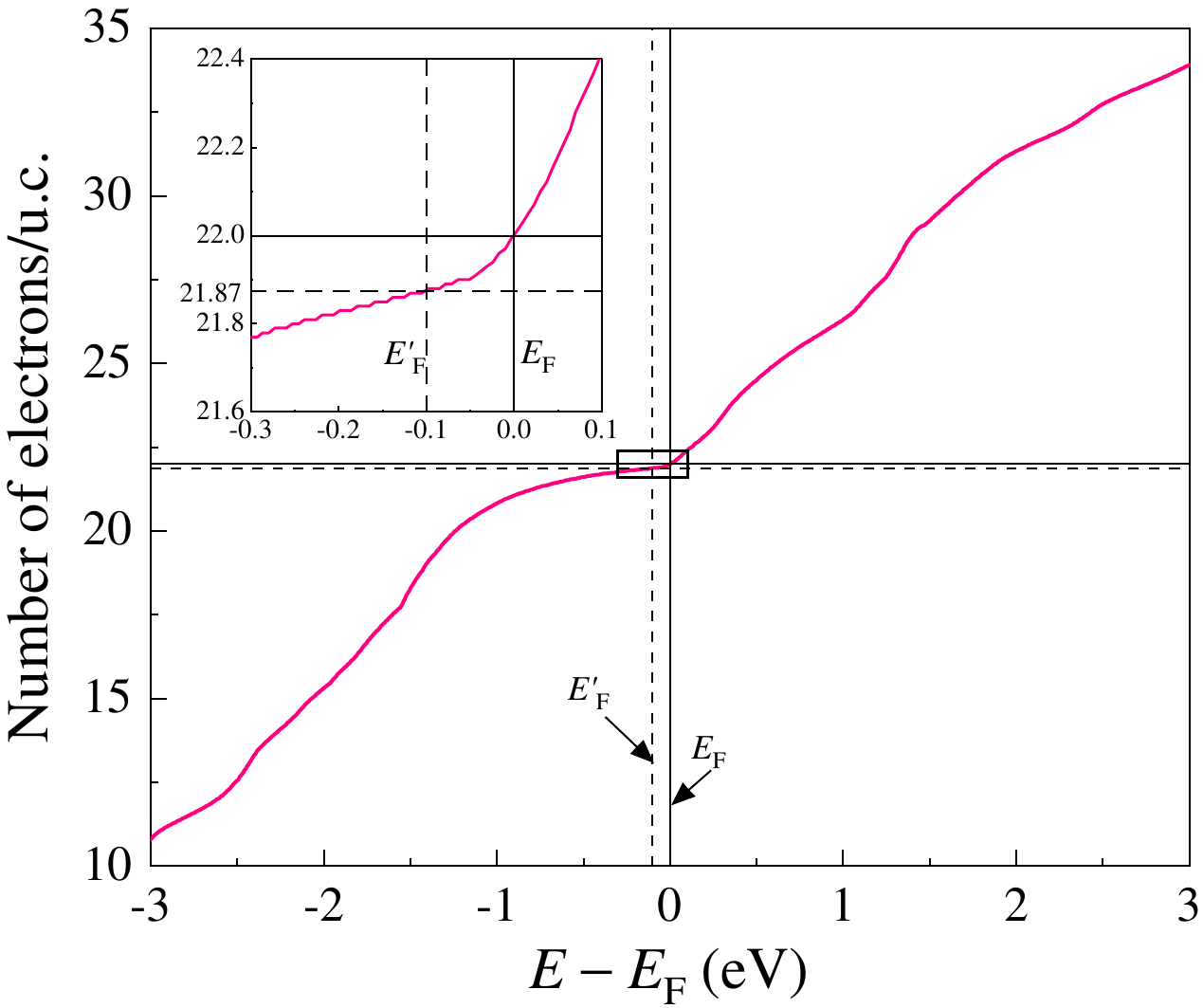}
\caption{Number of electrons per unit cell as a function of energy.}
\label{Fig s8}
\end{figure}

\section*{S7 \hspace{5mm}Seebeck coefficient of Cr$_{0.98}$V$_{0.02}$Sb}
To prove that a small amount of hole doping in CrSb destroy the DDCP, we have measured the in-plane Seebeck coefficient ($S_\mathrm{xx}$) of Cr$_{0.98}$V$_{0.02}$Sb. CrSb is n-type in the \textit{ab}-plane and thus has negative $S_\mathrm{xx}$. In Figure S10, we show that $S_\mathrm{xx}$ changes sign and becomes positive for Cr$_{0.98}$V$_{0.02}$Sb, thus destroying the DDCP. 
\begin{figure}[!h]
\renewcommand{\thefigure}{S10}
\centering
\includegraphics[width=0.65\columnwidth]{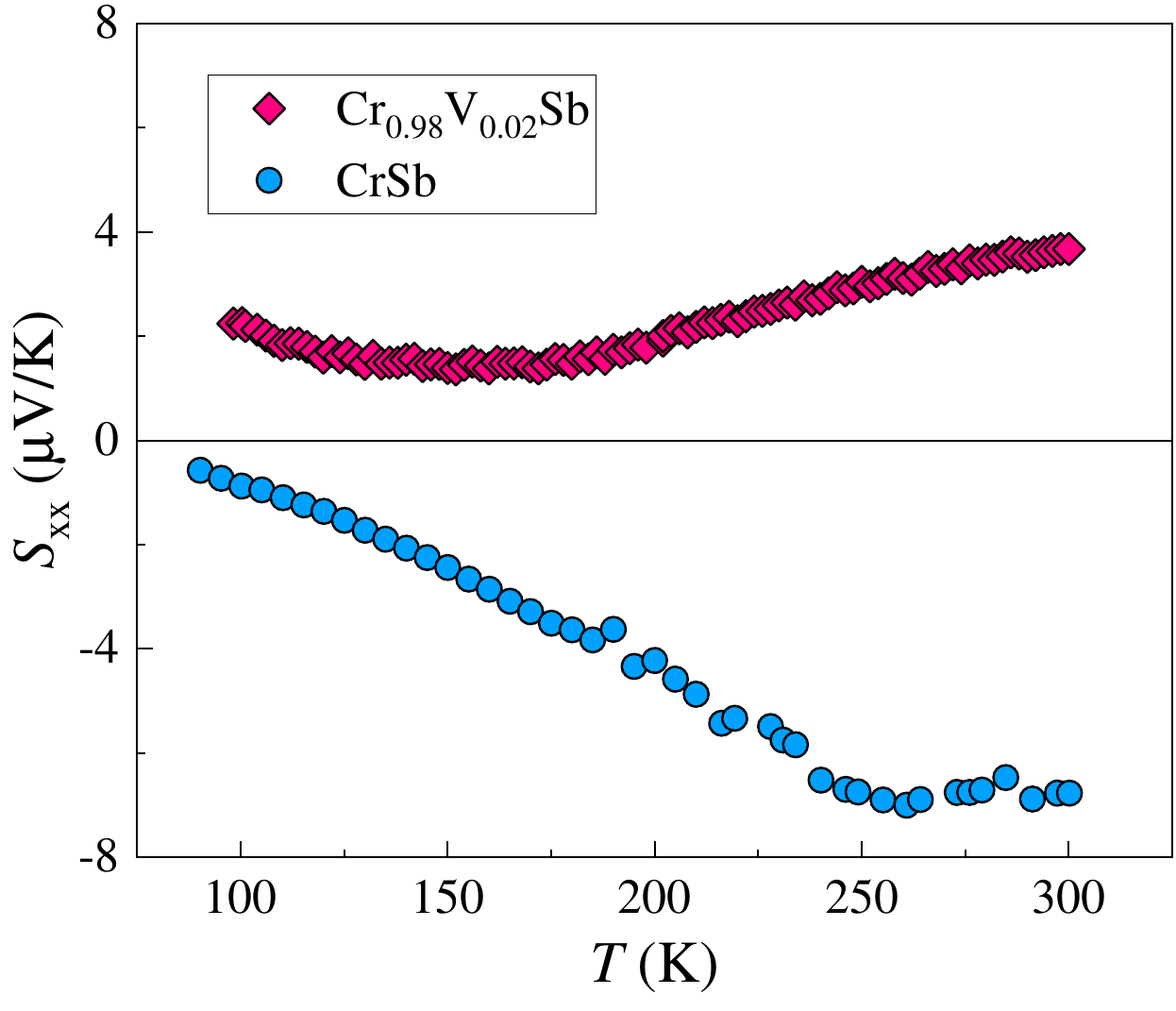}
\caption{In-plane seebeck coefficient ($S_\mathrm{xx}$) for CrSb and Cr$_{0.98}$V$_{0.02}$Sb, showing that the sign of $S_\mathrm{xx}$ has changed from negative to positive in Cr$_{0.98}$V$_{0.02}$Sb because of hole (vanadium) doping.}
\label{Fig s8}
\end{figure}


\input{Supporting_information.bbl}